# Agent-authored deposition recipes for X-ray multilayer mirrors: schema-bound LLM control of a magnetron sputtering system with reflectivity-verified outcomes

Oleksiy V. Penkov[a,b,*], Haoyu Fu[a,c], Yueying Zhan[a], Jingjing Peng[a], Pengyuan Wu[d], Jiong Jin[e]

[a] *ZJU-UIUC Institute, Zhejiang University, Haining, Zhejiang 314400, People's Republic of China*

[b] *Department of Mechanical Science and Engineering, University of Illinois Urbana-Champaign, Urbana, IL 61801, USA*

[c] *School of Materials Science & Engineering, Zhejiang University, Hangzhou 310058, China*

[d] *School of Mechanical and Electrical Engineering, Liuzhou Polytechnic University, No. 16 Guantang Avenue, Yufeng District, Liuzhou, Guangxi, China*

[e] *SVAC, Zhejiang Saiweike Optoelectronics Technology Co., Deqing, Huzhou, Zhejiang, China*

[*] Corresponding author. E-mail: openkov@illinois.edu

## Abstract

Periodic multilayer mirrors for X-rays require a layer period controlled to about 0.1 nm, set by deposition rates held in the laboratory's own notebook. A supervising model session gave a large language model agent a target multilayer specification and calibration constants derived from a laboratory archive. The agent authored recipes in the magnetron sputtering system's native recipe language and submitted them through a Model Context Protocol bridge of 17 tools in three tiers, with enforcement in the control application. The target was a 30-bilayer Ru/C multilayer with a period of 6.85 nm, a pair chosen because each period is fitted as two layers. The agent could read and write only its own working project, so it could alter neither the calibration record nor the measurements that scored it. We scored mirrors by X-ray reflectivity and fixed the 3% acceptance threshold on relative period difference before deposition. Across four scored depositions, the relative period difference fell from 20% on the calibration as given to 2.26% after one measured-feedback iteration and 0.79% after a second. We identified sixteen failure modes of the agent-facing interface. The result is bounded to the period, pair, and chamber observed.

## 1. Introduction

### 1.1 Background

Because no single material has a useful normal-incidence reflectivity for X-rays, periodic multilayer mirrors are used as optical elements in this range [1][2][3]. A multilayer mirror is a stack in which a pair of high- and low-atomic-number layers is repeated many times [4][5]. According to Bragg's condition, the layer period sets the wavelength and the angle at which the mirror reflects [5][6]. Individual layers are of the order of 1 nm.

Thus, period control must be maintained at the 0.1 nm level across the whole stack. The mirrors deposited in the present work had a period of 6.85 nm over 30 bilayers. Before starting a stack, the deposition rate of each material and the opening time of each shutter must be known. Because the rate depends on the target, sputtering power, working pressure, and source-to-substrate geometry, it is determined by calibration. A self-driving evaporation system reports the same difficulty and grows a sacrificial calibration layer on every sample [7]. Calibration data are specific to each laboratory and are kept in its notebook and in the operator's experience.

X-ray reflectivity gives an independent measurement of the deposited structure [8][9][10]. The Bragg peak positions give the period directly. A fit of the full curve gives the thickness, density, and interface width of each layer. Thus, each multilayer mirror can be accepted or rejected on its own reflectivity scan.

### 1.2 Related work

On a trapped-ion platform, an agent wrote native control code, ran it on live hardware behind a safety filter, and returned real physics results [11]. Frontier models designed the control system of a scanning helium microscope used for imaging and diffraction [12]. These models worked offline through web interfaces and were judged qualitatively against manual control. In another study, a model wrote code to acquire and plot photocurrents from a scanned sample [13]. No irreversible action was taken, and no quantitative metric was reported. Agents have been applied more widely [14], from autonomous chemical research [15][16] to atomic force microscopy, transmission electron microscopy and radio-frequency instrumentation [17][18][19]. Other work is schema-bound and runs only in software [20]. In one case, an agent ran a simulated ALD process, scored its own success, and could converge on undersaturated recipes [21]. None of these instruments is a deposition system.

Depositions have been automated from the other direction, with self-driving platforms for solution processing [22], effusion-cell evaporation [7], and magnetron sputtering [23]. Autonomous sputtering on live hardware has grown real films whose composition was steered by Bayesian optimization informed by optical emission spectroscopy [24]. In that work, the loop closed before the substrate shutter opened. A self-driving sputtering epitaxy system grew real films under Gaussian-process Bayesian optimization against the Urbach energy, an optical disorder metric [25]. In that system, film thickness was set open-loop to a nominal value from a prior rate calibration. A human who followed growth rules distilled from the optimizer's data reached an Urbach energy of 163 meV, compared with 182 meV for the optimizer itself [25]. Neither study used a language model. In both, the controlled quantity was composition or an optical metric rather than a structural length scale. When agents choose ALD conditions, the process is a physics-based simulation that lets conditions be explored without risking the reactor [26].

No language model has yet authored native recipe code for a real deposition campaign whose outcome was judged by a structural measurement against a human-authored reference. Where a process record is kept, it is the system's own datastore [23]. No such work reports an agent-driven campaign built on a laboratory's pre-existing record.

### 1.3 Contribution

In this study, we gave an LLM agent a target multilayer specification and calibration constants derived from a laboratory archive. The archive was read by a separate, supervising LLM session, which also passed the measurements to the agent. From these inputs, the agent authored deposition recipes in the native recipe language of a magnetron sputtering system. The recipes were submitted to the deposition tool through a Model Context Protocol bridge [27] with 17 tools in three capability tiers. Enforcement ran server-side in the control application. The agent's notebook account could read and write only its own working project. A person entered every measurement it was scored against, so no result reported here passed through the agent before it was recorded. X-ray reflectivity scored the deposited mirrors against a human-authored reference, with the metric and acceptance threshold fixed before deposition. We report all sixteen failure modes of the agent-facing interface identified during the work.

## 2. System

### 2.1 The deposition system

The work was carried out on a magnetron sputtering system with a vertical cylindrical chamber (Figure 1). The magnetrons are mounted on top, and the substrate holder below them moves from one magnetron to the next to deposit alternating layers. The exposure time of each layer is set by a shutter with a window. A stack of several hundred periods takes three to five hours to deposit, and the period must stay within 1% for that entire time, or about 0.02 to 0.04 nm for 2 to 4 nm layers.

Control is divided between two levels (Figure 2). A programmable logic controller runs the vacuum and gas subsystems, and the workstation can read their state but cannot command them. The workstation reaches the power, mechanics, and data-acquisition subsystems. Because a Windows workstation cannot maintain accurate time, critical timings, particularly the shutter exposure time, are delegated to programmable stepping-motor controllers. The workstation interprets a deposition recipe and sends these controllers short self-contained sequences, such as moving the holder or opening the shutter for a given time. A recipe is therefore authored two levels above the controllers on which the period tolerance is met.

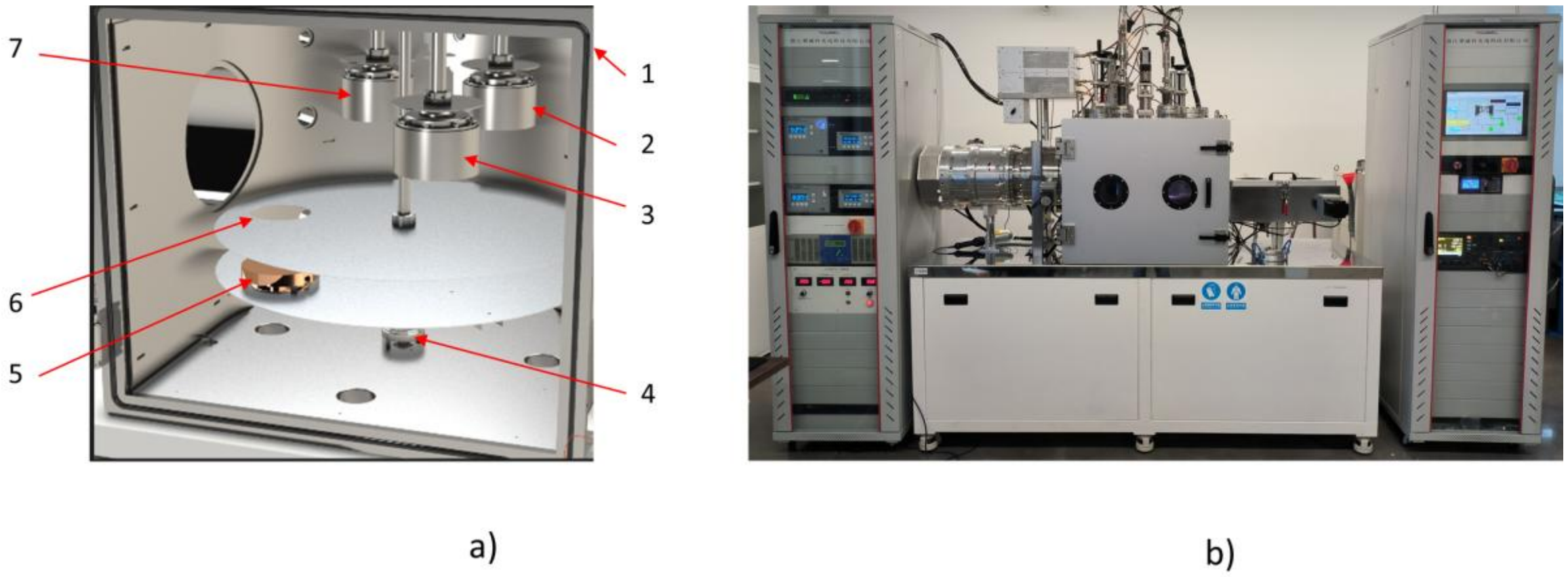

*Figure 1. The deposition system. (a) Cutaway of the vacuum chamber: 1, chamber; 2, 3 and 7, magnetrons; 4, robotic arm; 5, substrate holder; 6, shutter window. The substrate holder is moved beneath a chosen magnetron and the shutter window sets the exposure. (b) The installed system, with the power and control racks on either side of the chamber.*

The control software is APS System Control, part of the APS Software Suite (APS Technology) [28], which also includes a recipe editor and an emulator of the machine.

## 2.2 The recipe layer

The control application stores recipes in a procedural recipe language wrapped in XML (Section S1). A periodic multilayer of N bilayers is written once as a loop body, so the recipe does not grow with the number of periods. The laboratory's historical runs were recorded in the script format of the previous control software, so a reference recipe cannot be transcribed directly into the current language. The recipe executed for each deposition is retained with its run log in the laboratory notebook of Section 2.4.

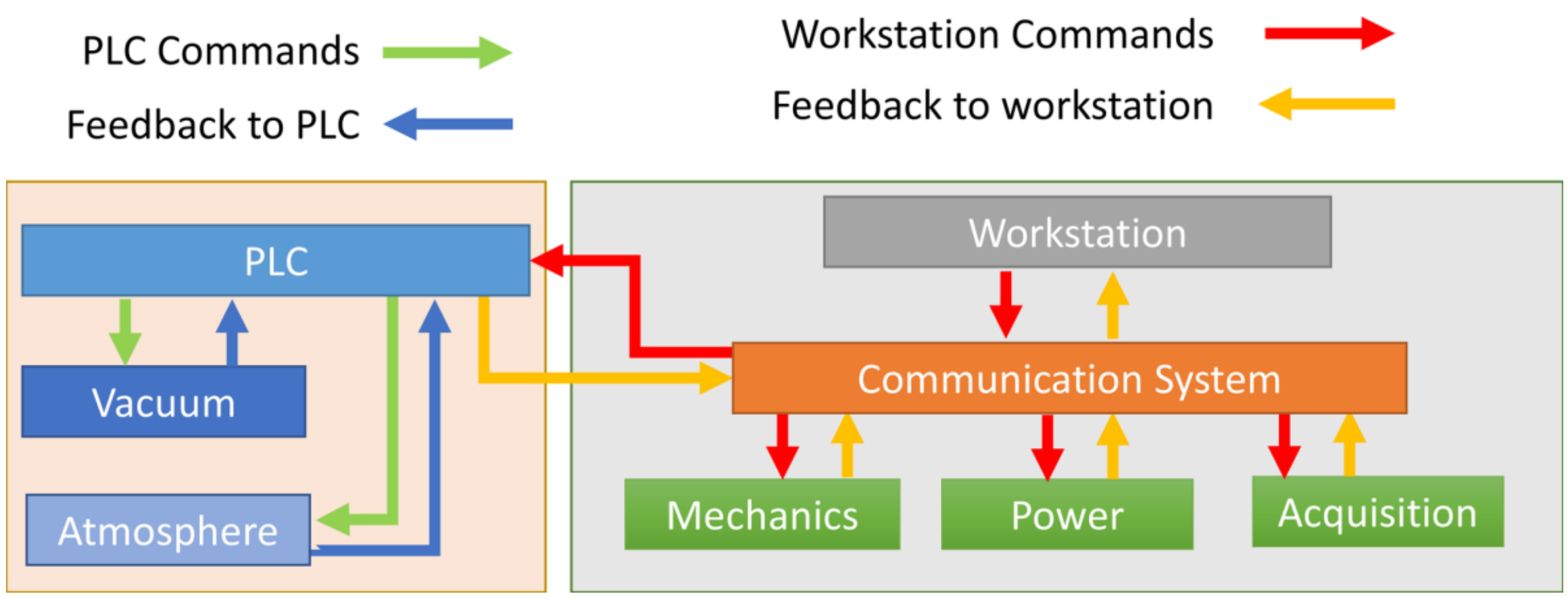

*Figure 2. Division of authority in the control system. The programmable logic controller carries the vacuum and atmosphere subsystems. The workstation reaches the mechanics, power and acquisition subsystems through the communication layer, and can read the vacuum state without commanding it. The atmosphere subsystem is reachable from either side. Arrow color distinguishes commands from feedback.*

### 2.3 The tool surface

The bridge is an MCP server with 17 tools (Figure 3). Each tool maps one-to-one onto an operation that the control application already exposes, so the bridge adds no capability of its own. The tools fall into three cumulative capability tiers: Read, Recipe and Manual (Section S2, Table S1). An agent granted the Read tier alone can inspect the machine but cannot upload or start a recipe.

Two tools serve authoring directly. The first returns the reference of the recipe language, including the device names and unit rules, from the machine at runtime. The reference therefore cannot drift from the machine it controls. The second validates a recipe without saving it, so authoring iterations do not use the hardware and their number can be counted. With no signatures loaded, the default-deny validator reports that it cannot vouch for a recipe rather than passing it.

### 2.4 The process record and its interface

The process record was kept in an electronic laboratory notebook, E-LabNotebook 3 (APS Technology) [29]. Each item carries files, recorded values, protocols, and signatures, and every creation, update, and deletion is logged with a user identity and a timestamp. A second MCP server, separate from the control bridge, exposes the notebook. Every write through it passes a server-side check that confines a user to projects that the user owns or has explicit write permission on.

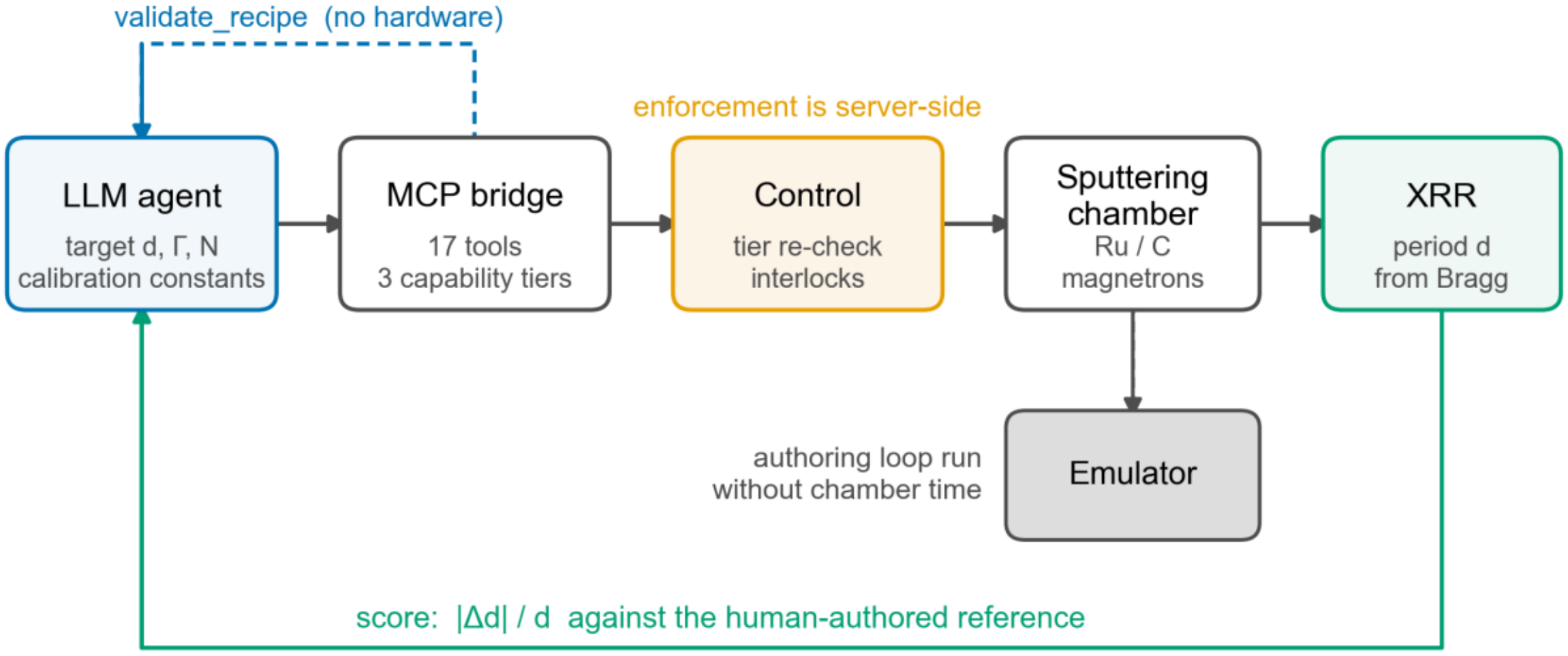

*Figure 3. Control path from the agent to the deposited sample. The agent authors a recipe in the native recipe language and submits it through the MCP bridge, which exposes 17 tools across three capability tiers. Enforcement of what the agent may do is held server-side in the control application, where the capability tier is re-checked and the interlocks are applied on the hardware thread. A recipe can be validated without reaching hardware, and the whole authoring loop can be run against the emulator. The deposited mirror is scored by X-ray reflectivity against a human-authored reference.*

The agent held one account that owned one working project, empty at the start of the campaign (Figure 4). Its credentials confined both reading and writing to that project. There, the agent recorded its depositions and specimens. From the second chamber run onward, it also attached the machine files of its runs, namely the process log, the data series, and the recipe as run. The measurement files, namely the profiler result, the reflectivity scan and its fit, were attached by a person, as was every value the agent was scored against. The agent therefore could not alter, select, or fabricate the measurements that scored it, and every record it added carries its identity in the log.

## 3. A structural safety layer

The control application enforces what the agent may do. The agent has no safety logic of its own, and the bridge only forwards calls to the server. Thus, the boundary lies in the software that already controls the hardware. The agent's capability is bounded by construction rather than by instruction.

Five mechanisms operate server-side. The first is a license feature that controls whether the agent server starts. The second is an API key that carries a capability tier. A pair of deployment-wide ceilings forms the third and clamps every key's tier downward. Fourth, every incoming command is re-checked against the tier of the key that carries it. Fifth, the hardware thread enforces the interlock set. Section S7 gives the rules of each mechanism and the tools granted at each tier.

Because the tier schema is cumulative, a key also carries everything granted at the tiers below its own. The set granted by a key is fixed when the key is issued and is clamped further at start-up by the deployment ceilings. A call above the tier of its key fails at runtime and does not reach the hardware. Interlocks are enforced on the same path that serves the operator panels, so an interlock that stops an action from a panel also stops the equivalent call from the agent.

A published system for LLM-written control code on a trapped-ion platform protects its hardware differently, with authorization tokens bound to individual operations [11]. A token is issued automatically once the operation passes an isolated hardware simulation with per-device bounds checking. For sensitive operations, a human operator issues the token manually. In the arrangement described here, a tier-clamped key grants capability and the hardware interlocks bound it, with no token issued per operation.

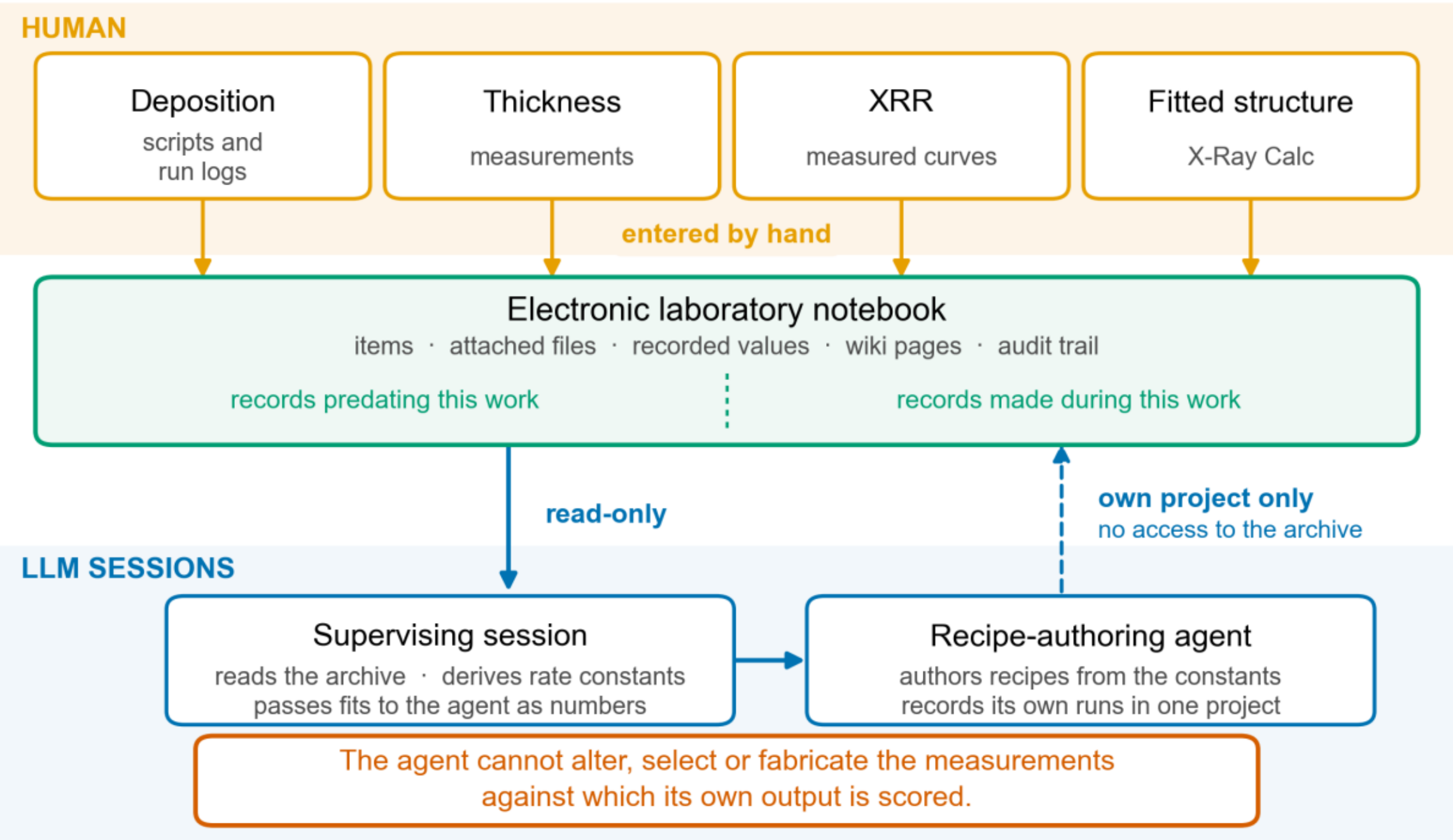

*Figure 4. The data path, and the division of labor on it. Every measurement in the laboratory notebook is entered by a person: the deposition scripts and run logs, the thickness measurements, the X-ray reflectivity curves, and the structures fitted from them in X-Ray Calc. The supervising session reads the archive through the notebook's Model Context Protocol server, which is separate from the control bridge of Figure 3, and gives the agent the derived calibration constants. The agent's own notebook access is confined to one working project, where it records its runs, so it can neither alter nor select the measurements against which its output is scored. Every write is checked on the server against project ownership or an explicit write permission. No one has tried to break that boundary on purpose.*

The safety layer described here has not been tested against deliberate attempts to break it. In a red-team campaign against its own deployed filter, the trapped-ion work ran 1932 attack scripts and approximately 250 harness-level tests [11]. A control set of 631 legitimate calibration scripts was approved without a single false rejection. The same campaign found the static stage of that filter to be evadable. No equivalent campaign was run here. Thus, we report the five mechanisms as documented and as they read in the source, with no claim that they resist a determined attempt to defeat them.

The safety layer constrains the machine but has no model of the deposition process. A recipe that is physically valid and passes every check can still produce a poor mirror. For this reason, the deposited structure is judged outside the safety layer. Several process rules are project policy rather than machine interlocks, such as the power-supply mode, ramp-rate limits on target power, bounds on deposited thickness and run duration, and human authorization for a target change or a new material pair. The written prescription holds these rules, and the operator applies them.

## 4. Hazards of an agent-facing hardware API

Four hazards of the agent-facing interface apply to any hardware API of this kind. First, every mutating tool returns as soon as the control application accepts the request, and a successful return carries no information about its effect. Interlock denials, alarm trips, and the failure protocol are asynchronous and appear only in the event log. The required pattern is verify-after-command, in which a second call reads the device state or the recipe status.

Second, a mutating command interrupted by a lost connection leaves an outcome that the response cannot resolve. After a reconnect, the bridge retries only read commands, because a mutating request may already have executed before its response was lost. The caller therefore has to re-query the state rather than assume that the command failed.

Third, failures can be silent or slow. The argument shape of the manual device command differs between device kinds, and one shape was accepted, logged nowhere, and discarded, so the caller believed a power supply was energized while it stayed off (Section S7). An abort triggers the full failure-protocol shutdown, which took on the order of 30 s, so an abort returns long before the machine is safe. A blocking operator dialog raised by a recipe survives the abort and can stall that shutdown until a person dismisses it.

Fourth, the state an agent reads can be stale, and the interface itself can be absent, with neither condition announced. The active-alarm list stays populated until an operator resets it, so it shows only that an alarm occurred since the last reset. A disabled deployment, a missing key, and an absent license feature all look the same at the client, as a server that cannot be reached.

## 5. Reading a real process record

The laboratory archive is the notebook of Section 2.4. It holds the Ru/C series deposited before this work, with its scripts, run logs, fit files, recorded values, and a wiki page of deposition rates. The supervising session read the series using read-only calls. The recipe-authoring agent had no read access to it and received only the constants derived below (Section 6).

The rate page serving as the calibration record did not list the Ru/C series, so we derived both rates from individual fits (Table S2). Each Ru/C period is modeled as two layers, so the Ru thickness is a single fitted quantity and the period is the sum of two fitted layers. Such structure makes the reflectivity fit direct, and it was the reason the pair was chosen. The Ru rate was 0.04711 nm/s, with a spread of 1.5% over four runs. The C rate was 0.04030 nm/s, with a spread of 4.6% over six runs.

The archive had five kinds of defects. No fit was linked to its source file, and fits could be matched to files only by the coincidence of first-period values. For two samples, the recorded value came from a first fit, while later refits existed and disagreed. The sample tagged as the best had no script and no log, and its recipe survived only as a note that it repeated an earlier run. Its fit was the most uniform in the series, with the period varying by 1.1% through the stack, whereas the earlier run with the complete record varied by 2.6% and showed split Bragg peaks. One fit file was attached to the wrong sample, and no goodness-of-fit figure was stored for any fit. Thus, a file retrieved from the archive could not be assumed to be the file that produced the number stored beside it.

Despite these defects, three runs at identical timing gave mean periods of 6.854, 6.835 and 6.741 nm, a spread of 1.7%, and 0.3% over the two fits in which both layers were refined. Any reader of such an archive, human or model, has to decide which records are current and build a stable quantity out of fitted ones. A synthetic benchmark poses none of these steps, because its ground truth is defined by construction. Reading the archive was itself a task, performed here by the supervising session before the agent received its instructions. None of the agent-driven studies cited in Section 1.2 report this step.

## 6. Experimental design

Three recipes were planned. The reference was a 50-bilayer mirror deposited by a person in 2026 with the previous script format and control software (Section S6). The second was to be written by a person in the current format and software. The third was written by the agent from the calibration record alone and repeated four times. The two new recipes used 30 bilayers to save chamber time, which does not affect the result because the period is a per-bilayer quantity. The human-written recipe was the only control separating a failure of the agent from an effect of the format and software change. It was not deposited within this study, so that control is missing, and the agent's specimens were scored against the target period, which is within 0.06% of the reference mean.

The agent received a target period of 6.85 nm, a target Γ of 0.215, defined as the Ru thickness divided by the period, 30 bilayers, a 19.7 nm Ru adhesion layer beneath the first period, a glass substrate, and the calibration constants of Table S2. The adhesion layer is the reference buffer expressed as a thickness, so the agent still had to derive its deposition time. No cap was specified.

The agent's instructions were amended seven times, and each amendment is logged with its trigger in Section S6. Three amendments added process knowledge that the calibration record lacked, including the vacuum and gas scheme and a required check that both discharges are struck before any material reaches the substrate. One corrected a note derived from the emulator that was wrong for the real machine.

No reference script, recipe, or fit was supplied, and for the agent the withholding was enforced by the notebook server rather than promised. The agent's account was scoped to a single new working project, and the project holding the executed Ru/C scripts lay outside that scope. The software running the agent exposed only the two tool servers, so the notebook interface was its only route to any record (Section S6).

Well-posedness was checked before the task was issued. The constants give layer times of 31.26 s for Ru and 133.44 s for C, against 31 and 133 s in the reference recipe. These times give a period of 6.820 nm against the measured mean of 6.854 nm, a difference of 0.5%. Thus, a failure of the agent could not be ascribed to an ill-posed task.

Because reflectometry was measured outside the laboratory, a same-day screen of total thickness by optical profilometry was added before the chamber phase. A specimen that had delaminated, or whose total thickness fell more than 3% below the nominal 225.2 nm plus any cap, was not sent for reflectometry. The agent received the profiler result as a measurement and chose whether to send the specimen or to revise the recipe and deposit again the same day.

The emulator runs and the first chamber run used Claude Opus 5, and from the second chamber run onward the agent was Claude Fable 5.1, by the author's decision. Table 3 classifies the executions. The first chamber run does not count, because both attempts were aborted and nothing was deposited. The first execution of the second run ran on an empty holder, an operator omission outside the agent's observability, and also does not count. Before each chamber session, the operator cleaned and loaded the substrate.

The scoring was fixed before any deposition. The primary metric was the mean period from the reflectivity fit, which is set geometrically by the Bragg peak positions. The mean was chosen because the reference fit distributes layer thickness through the stack and has no single period, and any fit carrying such a distribution is reported with its first, last, and mean period. All curves were fitted in X-Ray Calc [9][30][10]. A recipe was scored a success when its relative period difference from the human-written recipe was at most 3%, with the same number of resolved Bragg orders. Because that recipe was not deposited, the difference was taken from the target period, and the number of resolved orders was compared with the eight of the reference. The spread of 1.7% over three reference-family runs makes 3% no tighter than the process reproduces. Γ, the range of the period through the stack, and the interface width were secondary metrics, and the interface width was not treated as reliable. The process metrics were the authoring iterations before validation passed, the validation errors, any capability-tier denials, the wall-clock and chamber time, and every human intervention (Section S6).

Authoring and execution were first exercised against an emulator of the deposition system, which required no chamber time. The instructions were developed there in five runs on an earlier target, whose records are withheld and will be reported separately, and in one run on the Ru/C target (Section 7.2). The four amendments from the emulator phase were in force for every chamber run.

## 7. Results

### 7.1 Reference characterization

The reference specimen had a mean period of 6.854 nm and a Γ of 0.215 over 50 periods. Its period varied by 1.1% through the stack, and its Bragg peaks were single up to the eighth order. A second specimen deposited from the same recipe, with a complete process record, had a mean period of 6.835 nm. From the third order upward, however, its Bragg peaks split into doublets corresponding to periods of 6.93 and 6.77 nm. Because the same recipe gave a uniform stack in the reference, the split was a process event rather than a property of the recipe. The process log showed no excursion in pressure, gas flow, or current, and no cause was verified.

Three runs at identical timing gave a spread of 1.7% in the mean period, and 0.3% over the two fits in which Ru was refined rather than held fixed. Thus, the 3% threshold of Section 6 was no tighter than the process reproduced.

## 7.2 Emulator phase

One Ru/C run was made against the emulator under the same instructions as the chamber runs, in an isolated session with a Recipe-tier key. The agent read the machine description, tested the validator, derived layer times of 31.26 s for Ru, 133.43 s for C, and 418.17 s for the adhesion layer, and validated a clean draft. It then stopped with a question about two recipe arguments. After the operator answered, it uploaded a recipe with no warnings and started it. The author stopped the run at period 24 of 30, after 1 h 24 min, to free time for the chamber runs, and the failure protocol returned the emulator to idle with no alarm. The run was not scored. The emulator shows only that a recipe runs to its end, and nothing about the film.

## 7.3 Chamber phase

Table 3. Chamber runs of the agent's recipes. The score is the relative difference between the mean fitted period and the 6.85 nm target, with a threshold of 3%. The human-written recipe was not deposited.

| Run | Outcome | Recipe origin | Mean period (nm) | Γ | Score | Result |
|---|---|---|---|---|---|---|
| 1 | not counted: two aborted attempts, no discharge | agent, calibration as given | n/a | n/a | n/a | not scored |
| 2, first execution | not counted: all steps run on an empty holder | agent, calibration as given | n/a | n/a | n/a | not scored |
| 2, second execution | complete | same recipe with a corrected ending | 8.254 | 0.144 | 20.5% | fail |
| 3 | complete | same recipe, second specimen | 8.191 | 0.146 | 19.6% | fail |
| 4 | complete | agent's correction from the two fits | 6.695 | 0.192 | 2.26% | pass |
| 5 | complete | agent's second correction | 6.796 | 0.208 | 0.79% | pass |
| 6 | complete | agent's choice: a single Ru film | single layer 12.82 | n/a | n/a | not scored |

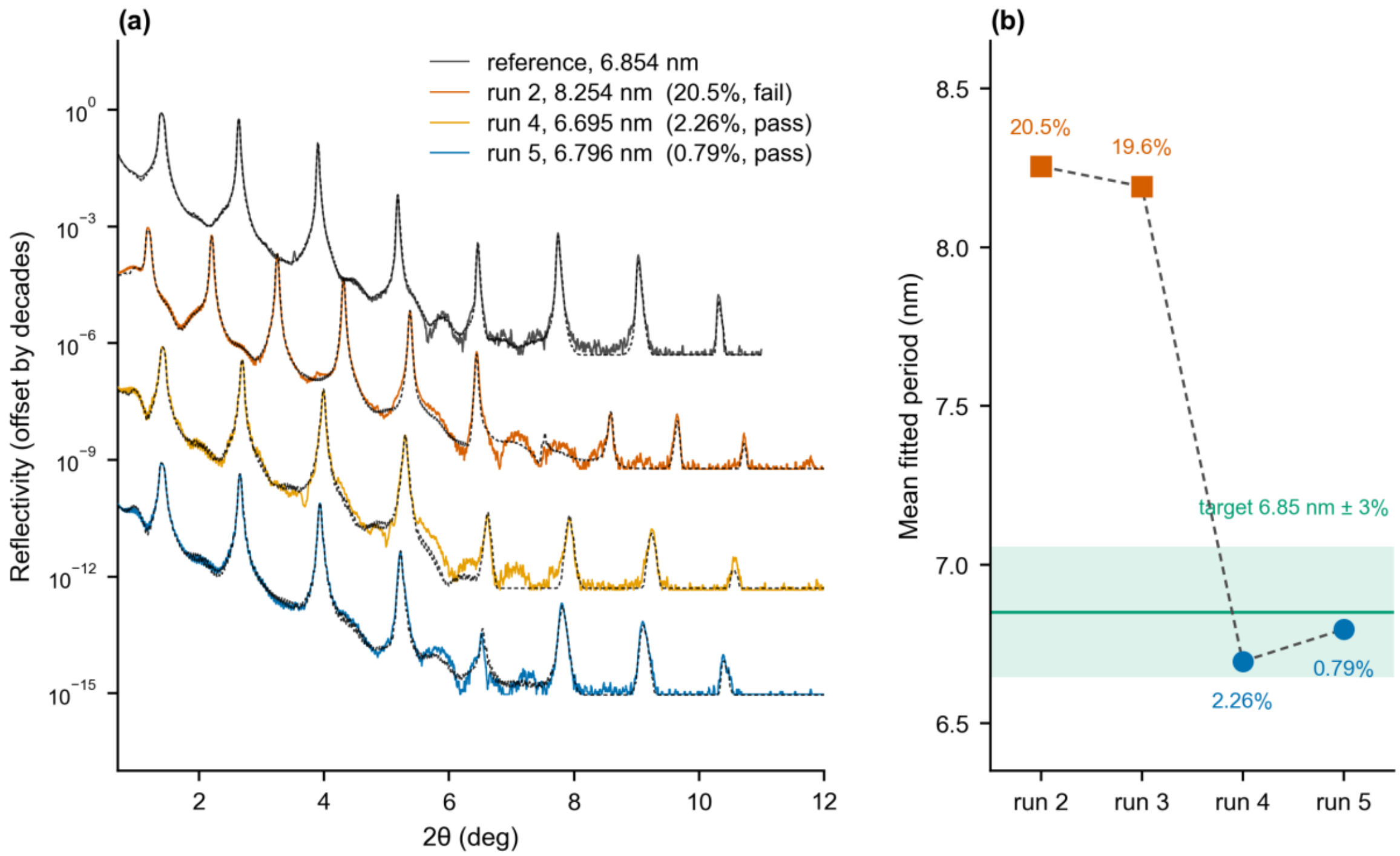

*Figure 5. The X-ray reflectivity result. (a) Measured curves (solid) with the structures fitted to them in X-Ray Calc (dashed), for the human-authored reference and for three agent-authored specimens, offset by three decades each for legibility and otherwise unaltered. Measured curves are drawn to 12 degrees, except the reference, whose scan ends at 11 degrees. (b) Mean fitted period of the four scored specimens against the 6.85 nm target and the pre-registered 3% threshold, with the score of each. Squares fail the threshold and circles pass it; the dashed line joins the runs in order and is a guide to the eye. Runs 2 and 3 carry one recipe on two specimens, so the campaign holds two corrections.*

The two specimens deposited from the calibration record as given had mean periods of 8.254 and 8.191 nm, 20.5% and 19.6% above the target, and both failed (Figure 5). The excess was carbon. The C layers of about 7.0 nm over a 133.43 s dwell gave a rate of 0.053 nm/s, 30% above the recorded 0.04030 nm/s, whereas the Ru layers of 1.19 nm fell short of their 1.473 nm nominal. The two specimens, deposited 3 h apart from one recipe, agreed to 0.8%. Thus, the error was systematic.

Given the two fits as numbers, the agent applied a rule it had stated in advance, a thickness $r(t - t_0)$ with a dead time $t_0$ of 6.07 s per deposit and rates of 0.04718 nm/s for Ru and 0.05516 nm/s for C. It moved the C dwell to 103.55 s and the Ru dwell to 37.29 s. Run 4 gave a mean period of 6.695 nm, a $\Gamma$ of 0.192, and a score of 2.26%, a pass on the first corrected run. The carbon rate came within 0.6% of the agent's value, and the remaining deficit of 2.3% was in Ru. The first-deposited period alone, 6.629 nm, would have scored 3.23%, but the score had been pre-registered on the mean.

Run 5 carried the agent's second correction, with Ru and C dwells of 43.3 and 103.0 s and the adhesion time unchanged. It gave a mean period of 6.796 nm, with 6.841 nm in the first-deposited period and 6.729 nm in the last, a Γ of 0.208, and a score of 0.79%. This result was within the band of 6.78 to 6.92 nm that the agent had set for itself, and the carbon rate was again within 0.7% of its model. The period decreased by 1.6% from the first to the last period, whereas in run 4 it had increased by 1.9%, and no cause was assigned. Both corrected specimens resolved eight Bragg orders, as the reference did. The two specimens with 8.2 nm periods resolved orders up to the tenth with the seventh absent, so neither matched the reference on that secondary criterion (Section 6).

With the buffer thickness free in the fits, the Ru adhesion layer fitted at 13.0 to 13.2 nm in all four specimens, against 19.7 nm nominal. Run 6, a single Ru film deposited for 423.65 s under the multilayer conditions, measured 12.82 nm, a rate of 0.0303 nm/s. This rate was 36% below the record and within 2.5% of the refitted buried buffers. The Γ of 0.19 to 0.21 against 0.215 reflected the same shortfall, so the mirrors were Ru-lean. The experiments were closed before the agent's revised adhesion time could be tested.

Every multilayer execution ran to its last step with no alarm, and the notebook record structure was correct from run 3 onward. The recorded API cost of the agent sessions was USD 42.28.

### 7.4 Failures and negative runs

A negative run is one that did not end with the machine in a correct state, and a defect is a fault in the software or its description that a run exposed. Table S3 in Section S4 lists the 16 defects.

Chamber run 1 was negative on both attempts, and nothing was deposited. The first attempt paused on a parameter name that the driver rejected, and the agent aborted. The second attempt struck no discharge, and the author ordered the abort. The first execution of run 2 ended with the throttle valve regulated nearly closed, because the instructions had named the gauge floor as the release target. That entry counts against the instructions rather than the agent. All later executions ended at base pressure with the valve released and no alarm.

Seven defects belong to the control application, its validator, its configuration, or a site recipe. The validator treated some characters inside comments as statements, and it did not check parameter names, so a wrong name failed only at runtime. The first pressure setpoint after a run did not engage the controller. This step paused every execution from the second execution of run 2 onward, five in all, and the agent's retry resolved it each time. A fault pause closed no valve and stopped no gas, because the machine profile defined no hold state. The emulator's valve model differed from the real controller, and a site cleaning recipe left one valve open. Two defects sit in the bridge. Its power-supply readback was stale, so the agent could not see the missing plasma in run 1, and one validation call hung without explanation. The remaining seven are on the record side and in the software that ran the agent. They include a mismatch between the served language description and the parser, stale reads from a long-lived notebook connection, and an agent process that ended when its copied credentials expired. A full run on an empty holder also completed normally, which was an operator omission outside the agent's observability.

No defect belongs to the agent. Its inaccuracies were confined to its reports, such as the claims that part of the adhesion layer had been deposited in run 1 and that the strike pressure had been reached in run 2. None of them entered a recipe.

## 8. Discussion

### 8.1 What the agent got right and wrong

Six things went right. Every uploaded recipe had passed the validator, and every multilayer execution ran every step. The fault pause at the first pressure wait occurred in five consecutive executions, and each time the agent chose to retry with no person involved. Its carbon correction was accurate to within 0.6% and 0.7% of its model in runs 4 and 5. Run 4 passed at 2.26% after one measured-feedback iteration, and run 5 passed at 0.79% after a second, inside the band of 6.78 to 6.92 nm and Γ of 0.205 to 0.225 that the agent had set for itself. In run 3, it declined to deposit again on profiler numbers while the reflectivity result of the first specimen was pending.

Five did not. The agent carried the Ru calibration into a 420 s adhesion layer, where the source delivered a third less, and it kept that time although one refit, and later three, had put the buffer at 13.1 nm. It read the depth profiles it was given as mean values only and never mentioned the variation of the period through the stack. In the author's view, run 6, a single Ru film, served the agent's rate model rather than the study, which needed another multilayer. When three uploads were refused because the control software was in local mode, it kept polling for remote mode instead of stopping as instructed. Its reports also carried the inaccuracies listed in Section 7.4, although none entered a recipe. None of these five are in the register, which lists defects rather than decisions.

### 8.2 Bounded autonomy

Authority over the recipe fell into three parts. The instructions fixed the target period, Γ, the bilayer count, the substrate, the working pressure, the gas flows, the adhesion thickness, and the calibration constants. The agent made the remaining choices, which are tabulated in Table S4 of Section S5. It derived the layer times from the constants in runs 1 to 3, and from a rate and dead-time model of its own, fitted to the reflectometry results, in runs 4 and 5. It chose the layer order, the discharge strike sequence and the valve settings, chose between abort and retry at each fault pause, and built the notebook record. It diagnosed problems with short test recipes rather than with direct device commands. With one deposition left, it chose a Ru calibration film over a second multilayer.

People made the other decisions. The author aborted the second attempt of run 1 upon seeing no discharge, changed the model from run 2 onward, released the throttle valve by hand after the empty execution, and restructured the notebook record after run 2. The operator switched the control software to remote mode, omitted the substrate in the first execution of run 2, and answered the agent's questions. The author refitted all scans, left the last deposition to the agent with the rule that a decline was final, and closed the experiments after run 6. The supervising session resumed run 3 after a credential failure, passed each fit to the agent as numbers only, and stopped two agent sessions, neither of which had reached the chamber.

The target, the constants, the pressure, and the material were fixed before the agent started. Its Recipe-tier credential allowed no device command outside a recipe. Substrate presence lay outside its observability, and a full execution completed on an empty holder. The measured-feedback step was exercised twice, in runs 4 and 5.

### 8.3 Where the safety layer would not have saved us

The safety layer of Section 3 is a boundary. It bounds what the agent may do and whether a recipe is well formed, but it holds no model of the film. Three events passed it. The first execution of the second run completed all its steps on an empty holder, because substrate presence lies outside the safety layer and the agent's observability. The Ru rate of 0.04711 nm/s, calibrated from multilayer fits with Ru layers of about 1.5 nm, was applied to a 19.7 nm adhesion layer, and every fit with a free buffer returned about 13.1 nm. The recipe of the second and third runs validated with zero findings and produced a period 20% too long. The study design caught all three. The same-day screen found the bare substrate, and reflectometry found the period and the buffer.

### 8.4 Generalization

Two elements can be transferred to other laboratories. The first is authoring against a validated recipe language with server-side enforcement, so that the agent writes text bounded by the machine's own validator and tiers rather than commanding hardware. The second is the measured-feedback loop. When the error is systematic, as for two specimens from one recipe that were 0.8% apart and both 20% too long, one fit and one corrected recipe brought the period inside the threshold, and a second correction reduced the residual further. Specific to this laboratory are the recipe language and its tool tiers, the notebook, and a Ru source whose delivered rate had fallen a third below its record. The result is bounded to Ru/C at 6.85 nm, one chamber, and four scored specimens.

### 8.5 Versus Bayesian self-driving deposition

The self-driving deposition systems of Section 1 close a different loop. The sputtering platforms steer composition or an optical disorder metric by Bayesian optimization over a parameter space [24][25], and the evaporation system of Zheng et al. grows a sacrificial calibration layer on every sample and decides from it [7]. Each needs an objective, a parameter space, and a procedure already written. This study asks whether an agent can write that procedure from a laboratory's record, execute it under a capability check, and correct it from one measured result, with a layer period as the controlled quantity. The two approaches are complementary, and no comparison of outcomes is made.

## 9. Conclusions

In this study, an LLM agent authored native deposition recipes for a Ru/C multilayer from a laboratory's own calibration record, submitted them through a tool bridge with a safety layer, and had them executed on a real magnetron sputtering system. Four specimens were scored by X-ray reflectivity against a 3% threshold on the relative period difference, fixed before deposition. The difference was 20% on the calibration as given, 2.26% after one measured-feedback iteration in which the agent corrected its rates from two fitted periods, and 0.79% after a second. Sixteen failure modes of the agent-facing interface were exposed and are registered in Table S3. The archive itself was incomplete and partly stale, and reading it was a task of its own before any recipe could be written. The per-decision table of Section 8.2 records which choices were the agent's and which a person's. The result is bounded to this period, this material pair, this chamber, and four specimens.

## Supplementary information

The supplementary information contains the recipe language and the argument forms that failed during this work (S1), the 17 tools of the bridge (S2, Table S1), the calibration constants supplied to the agent (S3, Table S2), the register of defects exposed in the chamber phase (S4, Table S3), and the decisions of the chamber phase with the party that took each one (S5, Table S4). It also gives the pre-registration detail, namely the reference, the amendments to the agent's instructions, the withholding of the reference and the scoring rationale (S6), and the rules of the safety layer (S7).

## Conflicts of interest

O.V.P. is the founder and owner of APS Technology. The company develops and sells the APS Software Suite (APS System Control, APS System Builder, APS Recipe Editor and APS System Emulator), which is the control software evaluated in this work, and E-LabNotebook 3, the electronic laboratory notebook that holds the process record examined in Section 5. The two Model Context Protocol bridges described in Section 2 belong to the same two products. Y.Z. is a minority shareholder in APS Technology. J.J. is a director of SVAC (Zhejiang Saiweike Optoelectronics Technology Co.). SVAC built the deposition system used in this work and supplied three of its magnetron sources. The remaining authors declare no competing interests.

Three features of this manuscript were adopted in response to these interests. Section 3 describes how the capability check is constructed, and does not certify its behavior under deliberate misuse, because no one tried to break it on purpose. Section 4, Section 7.4, and Table S3 report the hazards and defects of the same commercial software rather than withholding them, and Section 5 reports the state of a laboratory record held in the same company's notebook. The acceptance criterion, a relative period difference of at most 3% from the human-written recipe, was fixed in Section 6 before any agent-authored recipe was run.

The deposition system provided the platform for the experiment rather than its object of study, and no comparison against equipment from other suppliers was made.

## Data availability

The X-ray reflectivity curves behind Figure 5 are available from the corresponding author as comma-separated files, one per specimen. Each file holds the measured reflectivity and the reflectivity calculated from the fitted structure on the same 2θ grid, for the human-authored reference, the four agent-authored multilayers of runs 2 to 5, and the Ru film of run 6. A further file gives the mean fitted period, score, and result of each scored run, the values of Table 3 and Figure 5b. The structures were fitted by the authors in X-Ray Calc, and the released curves are the contents of those fit files, unaltered.

The recipes the agent authored, its session records, the process logs, and the notebook items are held in the authors' laboratory record and are available from the corresponding author on reasonable request.

## Acknowledgements

This research received no external funding.

Generative artificial intelligence enters this work in two roles, and the two are separated here. The first is the object of study. The agent described in Sections 2 and 6 authored deposition recipes, and what it produced is a result of the work rather than a part of its writing. Section 6 names the models it ran under.

The second role is assistance in preparing the manuscript. Claude (Anthropic), operated through the Claude Code command-line interface, was used between 2026-08-30 and 2026-09-14, in the model versions Claude Opus 5 and Claude Fable 5.1, for four tasks. Sections of the text were drafted from fact sheets compiled by the authors and were then edited by the authors. Each drafting prompt named the fact sheet, the section to be produced, its word budget, and its acceptance criteria, and forbade any number or citation not present in that source. Reference metadata were checked against publisher landing pages, and six errors in the authors' earlier notes were corrected in this way. The X-ray reflectivity data were measured by the authors and were fitted by the authors in X-Ray Calc. The fitted thicknesses were read out of E-LabNotebook 3, the authors' own electronic laboratory notebook, through an interface written by the authors, and the rate constants of Table S2 were computed from them. Retrieval and tabulation of values already recorded were the assisted part, and every value was checked against its source file by an author. The scripts that draw the schematics of Figures 3 and 4 and the script that plots Figure 5 were written with the same assistance.

No text, number, citation, or figure entered the manuscript without author review. Figure 5 is plotted from the measured curves and the authors' fits, read out of the fit files without refitting, smoothing, interpolation, or any other alteration, and its panel (b) reproduces the periods of

Table 3. No image in this manuscript was generated or altered by an AI tool. Generative artificial intelligence was not used to produce, alter, or select experimental data, and it holds no authorship of this work. The authors take full responsibility for the content of the manuscript.

# Supplementary Material

*Agent-authored deposition recipes for X-ray multilayer mirrors: schema-bound LLM control of a magnetron sputtering system with reflectivity-verified outcomes*

Oleksiy V. Penkov[a,b,*], Haoyu Fu[a,c], Yueying Zhan[a], Jingjing Peng[a], Pengyuan Wu[d], Jiong Jin[e]

[a] *ZJU-UIUC Institute, Zhejiang University, Haining, Zhejiang 314400, People's Republic of China*

[b] *Department of Mechanical Science and Engineering, University of Illinois Urbana-Champaign, Urbana, IL 61801, USA*

[c] *School of Materials Science & Engineering, Zhejiang University, Hangzhou 310058, China*

[d] *School of Mechanical and Electrical Engineering, Liuzhou Polytechnic University, No. 16 Guantang Avenue, Yufeng District, Liuzhou, Guangxi, China*

[e] *SVAC, Zhejiang Saiweike Optoelectronics Technology Co., Deqing, Huzhou, Zhejiang, China*

[*] Corresponding author. E-mail: openkov@illinois.edu

## S1. The recipe language

This section summarizes the recipe language of Section 2.2 and the argument forms that failed during this work.

Commands fall into two groups. A fixed set, among them `loop`, `wait`, `pwait`, `message`, `check` and `abort`, is handled by the interpreter itself. Further operations, among them `set`, `stage`, and `MFC`, are supplied by a worker configuration. When both define the same name, the interpreter takes precedence.

The arity of `pwait` depends on the worker configuration. When the command is bound to a device that is both the throttle actuator and the pressure sensor, the form is `pwait <target> <timeout>`. When it is unbound, the sensor must also be named, as `pwait <sensor> <target> <timeout>`. An argument list that is correct under one configuration is therefore wrong under the other.

`set <supply> <parameter>` passes the parameter name to the driver unchanged. Only full names such as `Current`, `Frequency` and `DutyCycle` are accepted, abbreviated forms are ignored without an error, and the valid names differ between supply models. Because `set` resolves its target as a power supply, directing it at a stage fails at runtime, and a stage is moved with `stage <name> <position>` instead. Neither fault is visible when the recipe is written.

The validator requires devices to be named, although the runtime also accepts a numeric index.

## S2. The tool surface

Table S1 lists the 17 tools of the bridge summarized in Section 2.3. `fetch_run_files` copies the as-run recipe, the time series and the process-log window of one run to the agent's side and returns their paths and SHA-256 digests, so that the files can be attached to the notebook.

**Table S1.** The 17 tools of the MCP bridge, grouped by capability tier. Tiers are cumulative, so a Manual key also carries everything granted at Recipe and at Read.

| Tier | Tool | Purpose |
|---|---|---|
| Read | `describe_machine` | Identity of the connected system: config profile, chambers, version. |
| Read | `describe_hardware` | Device inventory and sputter sources, with flange-to-material and supply mapping. |
| Read | `describe_recipes` | Available recipes, user and embedded. |
| Read | `describe_dsl` | Recipe language reference for the installed version, including unit rules. |
| Read | `describe_interlocks` | Interlock rules, always enforced server-side. |
| Read | `describe_alarms` | Alarm rule definitions per chamber. |
| Read | `system_status` | Per-chamber state, recipe and deposition flags, active alarms, progress. |
| Read | `device_snapshot` | One snapshot of all cached device readings. |
| Read | `tail_log` | Most recent event-log entries, including interlock denials. |
| Read | `recipe_status` | Running flag, current step, elapsed and remaining time. |
| Read | `list_runs` | Most recent runs with identifier, recipe, times, outcome and available files. |
| Read | `fetch_run_files` | Copy a run's as-run recipe, time series and log to the agent's side. |
| Recipe | `validate_recipe` | Validate content without saving; returns line-numbered errors. |
| Recipe | `upload_recipe` | Validate then save a recipe onto the machine. |
| Recipe | `start_recipe` | Start a recipe by name, with optional parameter overrides. |
| Recipe | `abort_recipe` | Abort the running recipe; triggers the failure-protocol shutdown. |
| Manual | `device_command` | Direct manual device command, same path as the operator panels. |

## S3. Calibration constants

Table S2 gives the deposition rate constants supplied to the agent. They were derived in Section 5 from individual XRR fits, because the notebook held no rate summary for this material pair.

**Table S2.** Deposition rate constants derived from XRR fits and supplied to the agent. Spread is the full range across the runs, divided by their mean.

| Material | Current (mA) | Rate (nm/s) | Spread (%) | Runs |
|---|---|---|---|---|
| Ru | 50 | 0.04711 | 1.5 | 4 |
| C | 300 | 0.04030 | 4.6 | 6 |

## S4. Failures register of the chamber phase

Section 7.4 of the main text summarizes this register.

Table S3. Defects exposed by the agent's chamber runs, with the component at fault and the status at the close of the experiments.

| # | Defect | Where | Exposed by | Status |
|---|---|---|---|---|
| 1 | The validator treats `;`, `/` or `-` inside a comment as a statement | control software, validator | run 1 | open; the recipe is validated only after the comment text is removed |
| 2 | The validator does not check `set` parameter names, so a wrong name fails at runtime | control software, validator | run 1; checked again in runs 2 and 3 | open |
| 3 | The served language description gives `shutter` an open or close argument, but the parser takes a position | bridge, language description | run 2 | open |
| 4 | `pwait` returns on the first sample past its target, so a gas-inrush transient satisfied the strike-pressure wait | control software | run 2, first execution | documented behavior; the named strike pressure was never reached |
| 5 | The first pressure setpoint after a run does not engage the controller until a second write | control software or pressure controller | runs 2 (second execution) to 6 | open; the agent's retry resolved it in all five executions |
| 6 | A release wait on the gauge floor leaves the real throttle valve nearly closed, whereas the emulator opens it fully | emulator model, carried into the instructions | run 2, first execution | fixed in the instructions by amendment 6; the emulator model still |

| | | | | differs |
|---|---|---|---|---|
| 7 | The machine profile defines no hold state, so a fault pause closes no valve and stops no gas | configuration | runs 2 and 3 | open |
| 8 | The site cleaning recipe leaves the ion-source gas inlet open | site recipe | run 2 | open; closed by hand |
| 9 | The power-supply readback stayed stale while the supplies delivered 0 V, so the agent could not see the missing plasma | bridge, control readback | run 1 | open; the cause of the missing discharge is not established |
| 10 | File upload through a project-scoped key creates file rows that no item lists | notebook server | run 2 | open; files re-attached by the supervising session |
| 11 | Agent-started runs get no signed conditions protocol, and the machine data series does not match the notebook's log layout | control software, notebook | run 2 | open |
| 12 | A long-lived notebook connection serves item text from before a desktop correction | notebook MCP server | runs 2, 3 and 5 | open |
| 13 | The software running the agent saved a large tool result to a file behind a short preview while the agent's file tools were disabled | agent software | run 2 | open; the agent proceeded on the preview |
| 14 | The agent process ended mid-run when its copied credentials lost a refresh race with the supervising session's | agent software | run 3 | open; resumed, recipe unaffected |
| 15 | A full run on an empty holder completed normally, because substrate presence is not sensed | machine, operator | run 2, first execution | operator omission, not a software defect |
| 16 | A validation call hung in the agent's bridge process while the same recipe validated elsewhere | bridge, client side | run 1 | open, unexplained |

## S5. Decisions of the chamber phase

Section 8.2 of the main text summarizes who decided what.

Table S4. Decisions of the chamber phase, the party that took each one, its basis, and the runs in which it applied.

| **Decision** | **Taken by** | **Basis** | **Runs** |
|---|---|---|---|
| Period, Γ, bilayer count, substrate, working pressure and gas flows | instructions | target and process conditions of Section 6 | 1–5 |
| Ru adhesion layer thickness | instructions | author's addition to the instructions | 1–6 |
| Calibration constants | instructions | calibration record of the pair | 1–3 |
| Amendments 1–4: throttle note, polling rule, vacuum and gas scheme, cleaned substrate | instructions | emulator phase | 1–6 |
| Amendment 5: source assignment, strike with the shutter open, discharge check before deposition | instructions | run 1 | 2–6 |
| Amendment 6: release wait targets zero | instructions | run 2, first execution | 2 (second execution) to 6 |
| Amendment 7: record hierarchy; the agent attaches the machine files | instructions | runs 1 and 2 | 3–6 |
| Layer times derived from the constants | agent | calibration constants and target | 1–3 |
| Layer order C then Ru, Ru-terminated, no cap | agent | target construction | 1–5 |
| Both supplies on for the whole stack, the shutter selecting the source | agent | recipe construction | 1–6 |
| Strike sequence, strike pressure and discharge check hold | agent | machine note of amendment 5 | 2–6 |
| Valves opened, unused valves closed | agent | vacuum and gas scheme of the instructions | 1–6 |
| Abort at the setpoint fault | agent | fault pause with nothing deposited | 1 |
| Retry at each fault pause | agent | log and status read at the pause | 2–6 |
| Specimen sent to reflectometry as it was | agent | profiler thickness given as a measurement | 2 (second execution) |
| Notebook record structure; machine files attached | agent | record protocol of the instructions | 1–6 |
| Diagnosis by short test recipes rather than device commands | agent | credential carries recipe tools only | 1–3 |
| Deposit again with its own rate correction | agent | the two reflectometry fits; its rule stated in run 3 | 4 |
| Rate and dead-time model; carbon rate 37% above the calibration; layer times 423.65, 103.55 and 37.29 s | agent | the two reflectometry fits | 4 |
| Pass window and prediction stated before the run | agent | its own model | 4, 5 |
| Adhesion time kept at the calibration value | agent | fitted adhesion thickness judged unreliable | 4, 5 |

| Second correction after run 4 passed, tRu 43.3 s and tC 103.0 s | agent | run 4 refit, below the target in $d$ and $\Gamma$ | 5 |
|---|---|---|---|
| Ru calibration film instead of a second multilayer | agent | three refits at 13.1 nm against the 19.7 nm expected | 6 |
| Abort of the second attempt | author | no discharge seen at the machine | 1 |
| Control software switched to remote mode | operator | upload refused in local mode | 1 |
| Model changed to Claude Fable 5.1 | author | author's decision | 2–6 |
| Substrate omitted; re-run called | operator | operator omission | 2 (first execution) |
| Throttle valve released by hand | author | valve left nearly closed | 2 (first execution) |
| Notebook record restructured | author | the laboratory's record hierarchy | 2 |
| Answers to the agent's questions | operator | control mode, message format, measurement files, profiler numbers | 1, 2 |
| Run 4 goes ahead on the agent's correction | author | both fits failed the pre-registered score | 4 |
| Reflectometry fits given to the agent as numbers only | supervising session | measured-feedback protocol | 4–6 |
| Specimen sent to reflectometry without a profiler screen | author | profiler judged not precise enough | 4 |
| Refit with the buffer free | author | first fit's buffer at a bound or in a local minimum | 2–4 |
| Profiler screen treated as coarse; reflectometry decides | author | profiler read 0.91–0.92 of the reflectometry total | 5 |
| Last deposition left to the agent; a decline is final | author | reflectometry of run 5 due the next day | 6 |
| Run resumed after the credential failure | supervising session | agent process ended | 3 |
| Two agent sessions stopped before any upload | supervising session | a superseded fit; polling instead of stopping | 5, 6 |
| Experiments closed | author | author's decision | after 6 |

## S6. Pre-registration detail

### S6.1 The reference

The reference set was the Ru/C series deposited in April and May 2026. Three runs were deposited at identical timing, with Ru for 31 s and C for 133 s per period. Their mean periods spanned 1.7%, and the two fits in which Ru was refined spanned 0.3%. The run with the most uniform fit was chosen as the reference, although it had no script or log. Its recipe was attested by a note that it repeated an earlier run and by an identical script deposited the same day. The run with a complete record was rejected, because its period ranged over 2.6% through the stack, from 6.935 to 6.759 nm, and its Bragg peaks split from the third order upward. In the reference fit, the period ranged over 1.1%, from 6.873 nm in the first period to 6.802 nm in the fiftieth, with a mean of 6.854 nm, and the Bragg peaks were single. The same recipe thus gave a uniform stack in one run and a drifting stack in the other, so the drift was a process event. A fourth fit, filed under one sample but titled for another, was excluded because its sample could not be retrieved from the archive. Each period of a Ru/C fit is modeled as two layers, so the period, the Ru thickness, and $\Gamma$ are single-valued fitted quantities.

The reference recipe specified 50 bilayers on a glass substrate, with C at 300 mA for 133 s and Ru at 50 mA for 31 s per layer, and C deposited first within each period. The working pressure was 2 mTorr, with two gas flows of 40 sccm each, and both supplies were pulsed at 50 kHz and 60% duty. A Ru buffer was deposited for 419 s before the first period, and a C cap for 33 s after the last.

### S6.2 Amendments to the instructions

The instructions were amended seven times. Amendments 1 and 2, made in the emulator phase, added a note on the throttle valve at the end of a run and a rule to poll the run status at intervals of 120 s or more. Amendment 3 described the vacuum and gas scheme, because every emulator run had to work out the gas routing for itself. Amendment 4 stated that the operator cleans and loads the substrate at base pressure before each session, because the ion-source current is set by hand and the recipe language cannot wait for a person. Amendment 5 followed the first chamber run, in which no plasma appeared for a reason that was not established. It added a machine note on the source assignment and the discharge strike, and it required a recipe check that both discharges are lit before any material reaches the substrate. Amendment 6 set the release wait at the end of a recipe to zero rather than the gauge floor, correcting a note taken from the emulator's controller model. Amendment 7, made before the third chamber run, stated the record hierarchy: one deposition record, one specimen beneath it, and the two measurements beneath the specimen. The agent had inverted that hierarchy in both earlier runs. Amendments 3 to 5 added process knowledge that the calibration record lacked, and amendment 7 added none.

### S6.3 The valve-label defect

While preparing the gas scheme for amendment 3, the author found that the machine configuration labeled the argon valve as nitrogen and the nitrogen valve as argon, in every profile, including that of the real machine. The emulator followed the same file. Every earlier emulator recipe was therefore correct against the description and wrong against the machine, and the safety layer had no means to notice. The labels were corrected before the Ru/C emulator run and the chamber runs.

### S6.4 Withholding the reference

For the agent's runs, its notebook credential was scoped to one new working project, and the notebook server checks that scope in every read tool. The project holding the executed Ru/C scripts lay outside it. A request outside the scope returns the same refusal as a request for a record that does not exist. The scope was tested before the campaign through the agent's own credential. The project listing returned only the working project, searches for the material and for the sample codes of the reference runs returned nothing, and direct requests for the reference items were refused. The test was repeated after the notebook server was updated to allow the agent to write, and the reference project was again refused for reading and writing. The software running the agent exposed only the two tool servers, with no file, shell, or web tool.

### S6.5 Well-posedness

The layer times follow from the constants. For Ru, $0.215 \times 6.85$ nm / 0.04711 nm/s gives 31.26 s, and for C, $0.785 \times 6.85$ nm / 0.04030 nm/s gives 133.44 s, against 31 and 133 s in the reference recipe. Run forward, 31 and 133 s give 1.460 and 5.360 nm, a period of 6.820 nm, which agrees with the measured mean of 6.854 nm to 0.5%. The check uses the mean period because the rates are means over all periods of each fit.

### S6.6 Scoring rationale

The fitting software can distribute layer thickness through a periodic stack to represent a deposition rate that drifted during the run [10]. The reference fit was made in this mode, with its period falling by 1.0% through the stack, so it has no single period. Because the Bragg condition then constrains the mean period, and the rates of Section 5 are means over all periods, the mean period was adopted in advance as the primary metric. The period range through the stack was a secondary metric, because it separated the two reference candidates, and $\Gamma$ was secondary as well. The interface width was not treated as reliable, because several fits returned a width larger than the layer it bounds.

### S6.7 The same-day screen

Reflectometry was measured outside the laboratory, with results returning the next day, whereas total thickness was measured at once by optical profilometry. A specimen that had delaminated, or whose total thickness fell more than 3% below its nominal 225.2 nm plus any cap, was not sent for reflectometry. A specimen that was too thick was sent, because its period could still be fitted. The agent received the profiler result as a measurement and decided whether to send the specimen or to revise the recipe and deposit again. The pre-registered score was taken on the first attempt that reached reflectometry, and a corrected attempt was reported as the outcome after one iteration of measured feedback.

## S7. The rules of the safety layer

### S7.1 Server-side mechanisms

The first mechanism is a license feature. Without it, the agent server does not start, no port is opened, and no error is shown. The second is the API key, which is stored only as a SHA-256 hash and carries a capability tier. The third is a pair of deployment-wide ceilings that clamp every key's tier downward. With the Manual tier disallowed, a Manual key acts as Recipe, and with Recipe also disallowed, as Read, so a key cannot exceed the ceiling set at the installation. The fourth re-checks every incoming command against the tier of its key, and a call above that tier fails without reaching the hardware. The fifth is the interlock set, enforced on the hardware thread on the same path as the operator panels. Interlocks cannot be bypassed through the API, and no manual device command is accepted while a recipe is running.

### S7.2 Tiers

Tiers are cumulative (Table S1). The Read tier grants 12 inspection tools, the Recipe tier adds the four tools that validate, upload, start and abort a recipe, and the Manual tier adds one tool for direct device commands. The tools granted by a key are fixed when the key is issued.

## S7.3 Transport

The bridge connects to the control application over TLS on the loopback address and pins the server certificate after the first connection. It holds no privilege beyond the API key it presents.

## S7.4 Device command shapes

The manual device command takes one of two argument shapes. Valves, mass-flow controllers, pumps and stages take a raw value, such as `[true]` for a valve or `[5.0]` for a flow setpoint. Power supplies, drives and controllers take a command name followed by values, such as `["SetCurrent", 250]` or `["OutputOn"]`. The recipe verbs `set` and `power` are not accepted in either shape. A mutating command is not retried after a lost connection, because a repeat could duplicate its effect, for example by toggling a valve twice.